\documentclass[reprint, amsmath, amssymb, aps, superscriptaddress, preprintnumbers]{revtex4-2}

\usepackage{graphicx} 
\usepackage{epstopdf}
\usepackage{dcolumn}
\usepackage{bm}
\usepackage[colorlinks=true,linktocpage=true,linkcolor=blue,citecolor=blue,urlcolor=blue]{hyperref}
\usepackage{textcomp}
\usepackage{color}
\usepackage{multirow}
\usepackage{tabularx}
\usepackage{tabu}

\newcounter{RSQ}

\begin{document}

\preprint{IFT-UAM/CSIC-24-82}

\title{Impact of NLO Weak SMEFT Corrections in $e^+e^- \rightarrow ZH$}

\def\BNL{High Energy Theory Group, Physics Department, 
    Brookhaven National Laboratory, Upton, NY 11973, USA}
    
\def\UR{Institute for Theoretical Physics, 
    University of Regensburg, 93040 Regensburg, Germany}
    
\def\IFT{Departamento de F\'{i}sica Te\'{o}rica and Instituto de F\'{i}sica Te\'{o}rica UAM/CSIC, Universidad Aut\'{o}noma de Madrid, Cantoblanco, 28049, Madrid, Spain}

\author{Konstantin Asteriadis}
\email{konstantin.asteriadis@ur.de}
\affiliation{\UR}

\author{Sally Dawson}
\email{dawson@bnl.gov}
\affiliation{\BNL}

\author{Pier Paolo Giardino}
\email{pier.giardino@uam.es}
\affiliation{\IFT}

\author{Robert Szafron}
\email{rszafron@bnl.gov}
\affiliation{\BNL}

\begin{abstract}
\noindent
We present results from a complete next-to-leading order (NLO) calculation of $e^+e^-\rightarrow ZH$ in the Standard Model Effective Field Theory (SMEFT) framework, including all contributions from dimension-6 operators.  At NLO, there are novel dependencies on CP violating parameters in the gauge sector, on modifications to the Higgs boson self-couplings,  on alterations to the top quark Yukawa couplings, and on 4-fermion operators involving the electron and the top quark, among others.  We show that including only the logarithms resulting from renormalization group scaling can produce misleading results, and further, we explicitly demonstrate the constraining power of combining measurements from different energy scales.
\end{abstract}

\maketitle

%%%%%%%%%%%%%%%%%%%%%%%%%%%%%%%%%%%%%%%%%%%%%%%%%%

\section{Introduction}
Future $e^+e^-$ colliders are designed to be precision machines capable of measurements at the percent level or better. This will allow discoveries of new interactions present only at the level of quantum corrections. One of the essential stages of the proposed FCC-ee~\cite{Bernardi:2022hny}  and CEPC~\cite{CEPCPhysicsStudyGroup:2022uwl} colliders for precision measurements of Higgs properties will be collisions at an energy of $\sqrt{s}=240~\rm GeV$, which optimizes the rate for associated $Z$ boson-Higgs production (Higgstrahlung), $e^+e^-\rightarrow ZH$. 
This process could also be probed at higher energies, such as $\sqrt{s}=365~\rm GeV$,  optimized for $t\bar{t}$ threshold physics.
In the Standard Model (SM) of particle interactions, the rate for Higgstrahlung has been computed to next-to-leading order in the electroweak interactions~\cite{Fleischer:1982af,Kniehl:1991hk,Denner:1992bc,Bondarenko:2018sgg}. Nearly complete calculations exist to next-to-next-to-leading order~\cite{Sun:2016bel,Gong:2016jys,Song:2021vru,Chen:2022mre,Freitas:2023iyx}. 
These results are the basis for sensitivity studies projecting future precision Higgs coupling limits, guiding the development of the physics program, and influencing the concept design of detectors. Moreover, we can expect the theoretical precision to increase.

The associated $ZH$ production process has the potential to open a window to physics Beyond the Standard Model (BSM). In particular, new physics effects on the couplings of the Higgs to fermions and gauge bosons can be consistently studied in the Standard Model Effective Field Theory (SMEFT) framework as an expansion in inverse powers of yet unknown physics at high mass scale. At leading order in the electroweak expansion, the $e^+e^-\rightarrow ZH$ cross-section depends on SMEFT interactions that have already been carefully probed at the LHC and with precision LEP measurements. Consequently, it is time to investigate the effects of new physics at the next-to-leading order in the electroweak expansion, where precision measurements hold potential for discovering new BSM phenomena in quantum fluctuations, as the cross-section acquires the dependence on interactions not probed at the leading order.

We report on a complete next-to-leading order electroweak calculation of $e^+e^-\rightarrow ZH$ in the SMEFT framework, including all dimension-6 operators and flavor structures that contribute. NLO QCD corrections in the dimension-6 SMEFT can be automated, but electroweak corrections must be done on a case-by-case basis and Drell Yan production is the only example of a 2- to- 2 scattering process where the NLO electroweak SMEFT calculations exist, making this calculation a significant advance~\cite{Dawson:2018dxp,Dawson:2021ofa}. In this letter, we show the sensitivity at a future $e^+e^-$ collider to SMEFT operators involving the Higgs boson tri-linear self coupling~\cite{McCullough:2013rea,Craig:2014una,Beneke:2014sba,Maltoni:2018ttu}, non-Standard Model top quark interactions, and CP violating Higgs-gauge boson couplings~\cite{Degrande:2023iob,ElFaham:2024uop}. Such anomalous Higgs interactions are predicted by many well-motivated BSM models, such as the 2 Higgs doublet model, the complex singlet model, or $Z^\prime$ models. 
Our results demonstrate that the contributions of different SMEFT operators are highly correlated, and limits based on single parameter fits can be highly misleading. This calls for developing a comprehensive strategy to disentangle various BSM effects at future colliders. In a companion paper, we present the details of the next-to-leading order calculation, including all dimension-6 SMEFT contributions and polarization~\cite{future}.

\begin{figure*}
	\centering	
    \includegraphics[width=.99\textwidth]{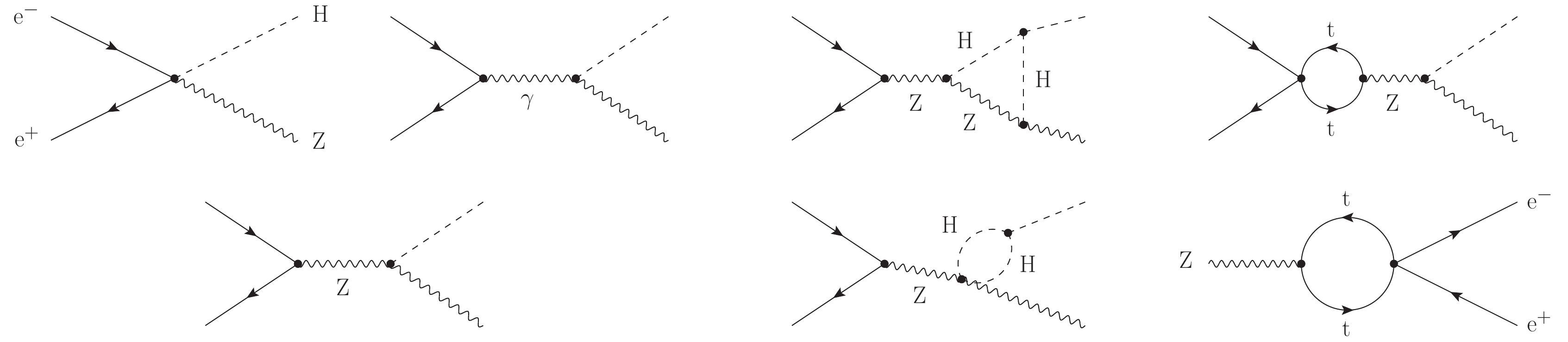} \\[7pt]
    \hspace{48pt}(a) \hspace{172pt} (b) \hspace{117pt} (c) \hspace{3pt}
	\caption{
        (a) LO contributions to $e^+e^-\rightarrow H Z$ in the dimension-6 SMEFT.  
        (b) Contributions sensitive to Higgs self-interactions arising from the operator $O_\phi$. 
        (c) Contributions from $O_{eu}[1,1,3,3]$.
    }
	\label{fig:diags}
\end{figure*}

\section{SMEFT Framework}
Deviations from the SM at high energies can be described in terms of an effective Lagrangian, which is an expansion around the SM,
\begin{equation}
    L=L_{\rm SM}+\sum_{i}\frac{C_i O_i}{\Lambda^2}+{\cal{O}}\biggl(\frac{1}{ \Lambda^4}\biggr) \, ,
    \end{equation}
where $O_i$ are ${\rm SU}(3)\times {\rm SU}(2) \times {\rm U}(1)$ invariant operators of  dimension-6 containing only SM fields and all BSM physics resides in the coefficient functions, $C_i$. 
The leading order (LO) dimension-6 SMEFT result arises from the diagrams shown in Fig.~\ref{fig:diags}~(a) and  we consistently neglect contributions of ${\cal{O}}({m_{\rm e}^2 / s})$ and ${\cal{O}}({1 / \Lambda^4})$.
We use the Warsaw basis and notation of~\cite{Grzadkowski:2010es}.

\section{Overview of the Calculation}

At tree level, the  cross section for $e^+e^-\rightarrow ZH$ depends on 7 SMEFT coefficients,
\begin{align}
\begin{split}
    &C_{\phi D}\, ,~C_{\phi \square}\, ,~C_{\phi WB}\, ,~C_{\phi W}\, ,~C_{\phi B}\, ,\\ 
    &C_{\phi e}[1,1]\, ,~C^+_{\phi l}[1,1] \equiv C_{\phi l}^{(1)}[1,1]+C_{\phi l}^{(3)}[1,1] \, , 
\end{split}
\end{align}
where the indices $[1,1]$ refer to the first generation fermions. 

Once higher order corrections are included, additional operators contribute. We compute the complete next-to-leading order (NLO) correction to $e^+e^-\rightarrow ZH$ in SMEFT,
including all dimension-6 one-loop  
effects up to ${\cal{O}}({g^2 v^2 / (16\pi^2\Lambda^2}))$ and the relevant real emission contributions. Here, we focus on the phenomenology of the contributions from the Higgs self-interactions, $C_\phi$, shown in Fig.~\ref{fig:diags}~(b), from operators that contain interactions with the top quark that are listed in Table~\ref{tab:top}, and from the CP violating operators, $C_{\phi {\widetilde{W}}B}$,~$C_{\phi {\widetilde{W}}}$,~$C_{\phi {\widetilde{B}}}$ and $C_{{\widetilde{W}}}$. Example diagrams involving 4-fermion  top quark-electron interactions are shown in Fig.~\ref{fig:diags}~(c). All these operators contribute to the cross-section for the first time at NLO. 

To calculate the relevant 1-loop diagrams, using the SMEFT Feynman rules~\cite{Dedes:2017zog}, we generate the amplitudes using FeynArts~\cite{Hahn:2000kx} and reduce them in terms of scalar Passarino-Veltman~\cite{Passarino:1978jh} integrals using FeynCalc~\cite{Shtabovenko:2023idz}. We chose a hybrid renormalization scheme, where SM quantities, particularly the masses of the gauge bosons $M_{\rm W}$, $M_{\rm Z}$ and the Fermi constant $G_{\mu}$ are renormalized on-shell, while the SMEFT Wilson coefficients are ${\overline{{\rm MS}}}$ quantities. Consequently, the cross section depends on the renormalization scale $\mu$.
This dependence can be deduced from a study of the one-loop  Renormalization Group Equations (RGE)~\cite{Jenkins:2013zja,Jenkins:2013wua,Alonso:2013hga}, and we verify that it agrees with our explicit calculation. Notably, the non-logarithmic corrections can be obtained only from the direct calculation. In particular, for $C_\phi$, $C_{u\phi}[3,3]$ and the CP violating operators, only this last contribution exists, and no information can be gained from the RGE.

For operators that contribute at the LO, infra-red divergences in the virtual contributions are treated using dimensional regularization, and we use dipole subtraction~\cite{Catani:1996vz,Denner:2019vbn} to regulate the real photon emission corrections. We perform the complete computation with a massless electron and then restore the leading dependence on the electron mass using collinear factorization and the fact that the electron mass plays the role of a collinear regulator~\cite{Bertone:2022ktl}. 

\section{Results}

We consider unpolarized electron-positron collisions at $240~{\rm GeV}$ and $365~{\rm GeV}$ center-of-mass energies. We assume $0.5\%$ accuracy for the measurement of the total cross section at $\sqrt{s}=240$ GeV ($1\%$ at $\sqrt{s}=365$ GeV) \cite{deBlas:2022ofj}.
Physical parameters are adapted from Ref.~\cite{Freitas:2023iyx}.
The Higgs boson is chosen to be stable with a mass of $m_{\rm H} = 125.1~{\rm GeV}$.
Vector boson masses are taken to be
$M_{\rm W}= 80.352~{\rm GeV}$ and $M_{\rm Z} = 91.1535~{\rm GeV}$, which include a shift to the measured values proportional to the experimental width, following Ref.~\cite{Freitas:2023iyx}, to account for finite width effects.
Relevant fermion masses are $m_{\rm e} = 0.511~{\rm MeV}$ and $m_{\rm t} = 172.76~{\rm GeV}$.
Weak couplings and the Higgs vacuum expectation value are derived from the weak boson masses and the Fermi constant $G_\mu = 1.1663787 \times 10^{-5}~{\rm GeV}^{-2}$.  The SMEFT scale is taken to be $\Lambda=1~{\rm TeV}$ for all numerical results.

Using these inputs, the leading order (LO) total cross sections at $\sqrt{s}=240~\rm GeV$ and $\sqrt{s}=365~\rm GeV$ are,
\begin{align}
\label{eq:sig2}
\begin{split}
    &\sigma_{\rm EFT,LO}^{(\sqrt{s} = 240~{\rm GeV})}({\rm fb}) = \sigma_{\rm SM, LO}^{(\sqrt{s} = 240~{\rm GeV})}   \\
    &\hspace{5pt}+25.3\, C_{\phi B} + 4.83\, C_{\phi D} + 29.0\, C_{\phi \square} + 133\, C_{\phi W}  \\ 
    &\hspace{5pt}+ 64.5 C_{\phi WB} - 177\, C_{\phi e}[1, 1] + 220\, C_{\phi l}^+[1, 1] \, ,
\end{split} \\[10pt]
\begin{split}
\label{eq:sig3}
    &\sigma_{\rm EFT,LO}^{(\sqrt{s} = 365~{\rm GeV})}({\rm fb}) = \sigma_{\rm SM, LO}^{(\sqrt{s} = 365~{\rm GeV})}   \\
    &\hspace{5pt}+21.9\, C_{\phi B} + 2.54\, C_{\phi D}+ 15.3\, C_{\phi \square} + 121\, C_{\phi W} \\ 
    &\hspace{5pt}+  55.6\, C_{\phi WB} - 216\, C_{\phi e}[1, 1] + 269\, C_{\phi l}^+[1, 1] \, , 
\end{split}
\end{align}
where $\sigma_{\rm LO,SM}^{(\sqrt{s} = 240~{\rm GeV})} = 239~{\rm fb}$ and $\sigma_{\rm LO,SM}^{(\sqrt{s} = 365~{\rm GeV})} = 117~{\rm fb}$ and all terms on the right hand sides of Eqs.~\eqref{eq:sig2}~and~\eqref{eq:sig3} are understood to be in fb.

At  NLO, the effects on the total cross section to ${\cal{O}}({1 / \Lambda^2})$ are parameterized as
\begin{align}
    \label{eq:sigpar}
    \frac{\sigma_{\rm NLO}}{\sigma_{\rm SM,NLO}} = 1 + \sum_i \frac{C_i(\mu)}{\Lambda^2} \bigg\{ \Delta_i + \bar{\Delta}_i \log{\frac{\mu^2}{s}}\bigg\} \, ,
\end{align}
and we show the size of these effects in Table~\ref{tab:top}. 
The NLO SM cross sections are~\cite{Freitas:2023iyx} $\sigma_{\rm SM,NLO}^{(\sqrt{s} = 240~{\rm GeV})}= 232~{\rm fb}$ and $\sigma_{\rm SM,NLO}^{(\sqrt{s} = 365~{\rm GeV})} = 113~{\rm fb}$.  

\begin{table}[t]
	\centering
 	\caption{Effects on the total cross section for $e^+e^- \rightarrow ZH$ from operators involving the Higgs self-interactions and from anomalous top quark interactions using the parameterization of Eq.~\ref{eq:sigpar}.	
  \label{tab:top}}	  
  \vspace{2pt}
  \begin{tabularx}{\columnwidth}{ccccc}
	\hline\hline
   & \multicolumn{2}{c}{$\sqrt{s} = 240~{\rm GeV}$}& \multicolumn{2}{c}{$\sqrt{s} = 365~{\rm GeV}$}
   \\
   & \hspace{6pt} $\Delta_i / \Lambda^2$    \hspace{6pt}
   & \hspace{6pt} $\bar\Delta_i /\Lambda^2$ \hspace{6pt}
   & \hspace{6pt}$\Delta_i / \Lambda^2$     \hspace{6pt}
   & \hspace{6pt}$\bar\Delta_i / \Lambda^2$ \hspace{6pt}
    \\
   \hline 
    $C_\phi$ & $\textrm{-}7.22 \hspace{-1pt}\cdot\hspace{-1pt} 10^{\textrm{-}3}$ & $0$ & $\textrm{-}1.00\hspace{-1pt}\cdot\hspace{-1pt} 10^{\textrm{-}3}$ & $0$\\
    \hline
    $C_{uW}[3,3]$ & $\textrm{-}1.63 \hspace{-1pt}\cdot\hspace{-1pt} 10^{\textrm{-}3}$ & $\hphantom{\textrm{-}}4.01 \hspace{-1pt}\cdot\hspace{-1pt} 10^{\textrm{-}3}$ & $\hphantom{\textrm{-}}3.36 \hspace{-1pt}\cdot\hspace{-1pt} 10^{\textrm{-}3}$ & $\hphantom{\textrm{-}}6.25 \hspace{-1pt}\cdot\hspace{-1pt} 10^{\textrm{-}3}$ \\
    $C_{uB}[3,3]$ & $\hphantom{\textrm{-}}0.15 \hspace{-1pt}\cdot\hspace{-1pt} 10^{\textrm{-}3}$ & $\textrm{-}2.22 \hspace{-1pt}\cdot\hspace{-1pt} 10^{\textrm{-}3}$ & $\textrm{-}2.96 \hspace{-1pt}\cdot\hspace{-1pt} 10^{\textrm{-}3}$ & $\textrm{-}3.20 \hspace{-1pt}\cdot\hspace{-1pt} 10^{\textrm{-}3}$\\
     $C_u\phi[3,3]$ & $\hphantom{\textrm{-}}0.32 \hspace{-1pt}\cdot\hspace{-1pt} 10^{\textrm{-}3}$ & $0$ & $\textrm{-}1.09 \hspace{-1pt}\cdot\hspace{-1pt} 10^{\textrm{-}3}$ & $0$\\
     $C_{\phi q}^{(1)}[3,3]$ & $\textrm{-}1.34 \hspace{-1pt}\cdot\hspace{-1pt} 10^{\textrm{-}3}$ & $\textrm{-}4.10 \hspace{-1pt}\cdot\hspace{-1pt} 10^{\textrm{-}3}$ & $\textrm{-}4.39 \hspace{-1pt}\cdot\hspace{-1pt} 10^{\textrm{-}3}$ & $\textrm{-}4.31 \hspace{-1pt}\cdot\hspace{-1pt} 10^{\textrm{-}3}$\\
    $C_{\phi q}^{(3)}[3,3]$ & $\hphantom{\textrm{-}}0.51 \hspace{-1pt}\cdot\hspace{-1pt} 10^{\textrm{-}3}$ & $\hphantom{\textrm{-}}4.12 \hspace{-1pt}\cdot\hspace{-1pt} 10^{\textrm{-}3}$ & $\hphantom{\textrm{-}}4.15 \hspace{-1pt}\cdot\hspace{-1pt} 10^{\textrm{-}4}$ & $\hphantom{\textrm{-}}7.58 \hspace{-1pt}\cdot\hspace{-1pt} 10^{\textrm{-}4}$\\
     $C_{\phi u}[3,3]$ & $\textrm{-}0.54 \hspace{-1pt}\cdot\hspace{-1pt} 10^{\textrm{-}3}$ & $\hphantom{\textrm{-}}3.49 \hspace{-1pt}\cdot\hspace{-1pt} 10^{\textrm{-}3}$ & $\hphantom{\textrm{-}}5.37 \hspace{-1pt}\cdot\hspace{-1pt} 10^{\textrm{-}3}$ & $\hphantom{\textrm{-}}3.11 \hspace{-1pt}\cdot\hspace{-1pt} 10^{\textrm{-}3}$\\
     \hline
     $C_{eu}[1,1,3,3]$ & $\hphantom{\textrm{-}}0.01 \hspace{-1pt}\cdot\hspace{-1pt} 10^{\textrm{-}3}$ & $\textrm{-}1.39 \hspace{-1pt}\cdot\hspace{-1pt} 10^{\textrm{-}2}$ & $\textrm{-}3.73 \hspace{-1pt}\cdot\hspace{-1pt} 10^{\textrm{-}2}$ & $\textrm{-}3.23 \hspace{-1pt}\cdot\hspace{-1pt} 10^{\textrm{-}2}$\\
     $C_{lu}[1,1,3,3]$ & $\textrm{-}0.02 \hspace{-1pt}\cdot\hspace{-1pt} 10^{\textrm{-}3}$ & $\hphantom{\textrm{-}}1.73 \hspace{-1pt}\cdot\hspace{-1pt} 10^{\textrm{-}2}$ & $\hphantom{\textrm{-}}4.64 \hspace{-1pt}\cdot\hspace{-1pt} 10^{\textrm{-}2}$ & $\hphantom{\textrm{-}}4.01 \hspace{-1pt}\cdot\hspace{-1pt} 10^{\textrm{-}2}$\\
     $C_{lq}^{(1)}[1,1,3,3]$ &  $\textrm{-}0.37 \hspace{-1pt}\cdot\hspace{-1pt} 10^{\textrm{-}2}$ &  $\textrm{-}1.80 \hspace{-1pt}\cdot\hspace{-1pt} 10^{\textrm{-}2}$ &  $\textrm{-}6.09 \hspace{-1pt}\cdot\hspace{-1pt} 10^{\textrm{-}2}$ &  $\textrm{-}4.18 \hspace{-1pt}\cdot\hspace{-1pt} 10^{\textrm{-}2}$\\
     $C_{lq}^{(3)}[1,1,3,3]$ & $\textrm{-}0.37 \hspace{-1pt}\cdot\hspace{-1pt} 10^{\textrm{-}2}$ & $\hphantom{\textrm{-}}1.29 \hspace{-1pt}\cdot\hspace{-1pt} 10^{\textrm{-}2}$ & $\hphantom{\textrm{-}}4.54 \hspace{-1pt}\cdot\hspace{-1pt} 10^{\textrm{-}2}$ & $\hphantom{\textrm{-}}3.29 \hspace{-1pt}\cdot\hspace{-1pt} 10^{\textrm{-}2}$\\
     $C_{qe}[3,3,1,1]$ & $\hphantom{\textrm{-}}0.30 \hspace{-1pt}\cdot\hspace{-1pt} 10^{\textrm{-}2}$ & $\hphantom{\textrm{-}}1.45 \hspace{-1pt}\cdot\hspace{-1pt} 10^{\textrm{-}2}$ & $\hphantom{\textrm{-}}4.90 \hspace{-1pt}\cdot\hspace{-1pt} 10^{\textrm{-}2}$ & $\hphantom{\textrm{-}}3.36 \hspace{-1pt}\cdot\hspace{-1pt} 10^{\textrm{-}2}$
     \\
    \hline\hline
\end{tabularx}
\end{table}

\begin{table}[t]
	\centering
 	\caption{Effect on total cross section for $e^+e^-\rightarrow ZH$ from CP violating operators. We take $\Lambda=1~{\rm TeV}$.	
  \label{tab:cp}}	  
  \vspace{2pt}
	\begin{tabularx}{\columnwidth}{ccc}
		\hline\hline
		& \multicolumn{2}{c}{$A_{{\rm CP},i} / C_{i}(\mu)$} \\
          \hspace{40pt}
        & \hspace{20pt} $\sqrt{s} = 240~{\rm GeV}$ \hspace{20pt}
        & \hspace{20pt} $\sqrt{s} = 365~{\rm GeV}$ \hspace{20pt} \\
        \hline 
        $C_{\phi {\widetilde{W}}B}$&$ 4.76 \hspace{-1pt}\cdot\hspace{-1pt} 10^{\textrm{-}3}$& $1.50 \hspace{-1pt}\cdot\hspace{-1pt} 10^{\textrm{-}2}$ \\
        $C_{\phi{\widetilde {B}}}$& $9.77 \hspace{-1pt}\cdot\hspace{-1pt} 10^{\textrm{-}4}$ & $1.94 \hspace{-1pt}\cdot\hspace{-1pt} 10^{\textrm{-}3}$ \\
        $C_{\phi{\widetilde {W}}}$ & $1.28 \hspace{-1pt}\cdot\hspace{-1pt} 10^{\textrm{-}2}$& $4.36 \hspace{-1pt}\cdot\hspace{-1pt} 10^{\textrm{-}2}$ \\
        $C_{{\widetilde {W}}}$ &$3.27 \hspace{-1pt}\cdot\hspace{-1pt} 10^{\textrm{-}3}$& $1.21 \hspace{-1pt}\cdot\hspace{-1pt} 10^{\textrm{-}2}$ \\
        \hline\hline
	\end{tabularx}
\end{table}

\begin{figure}
	\centering	
    \hspace{-15pt} \includegraphics[width=.39\textwidth]{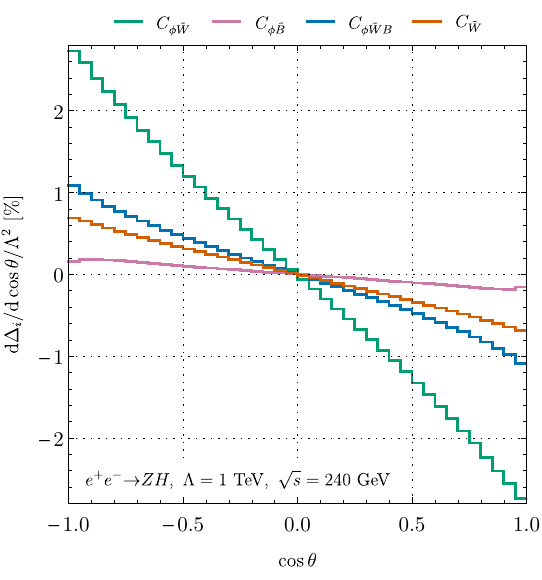} \\[5pt]
    \hspace{-15pt} \includegraphics[width=.38\textwidth]{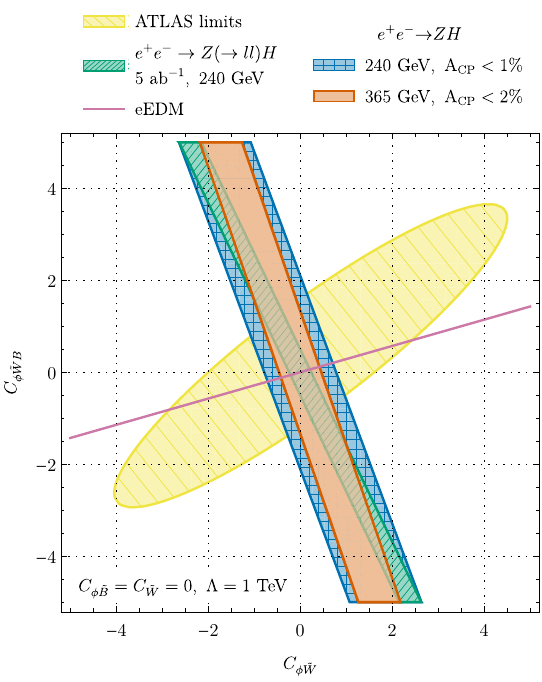} 
	\caption{
        Sensitivity to CP violating dimension-6 operators at NLO~(upper). 
        Sensitivity to CP violating dimension-6 operators. The current limits from ATLAS~\cite{ATLAS:2023mqy} and the electron EDM are also shown, along with a projection for angular observables at FCC-ee with $5\;~{\rm ab}^{-1}$~\cite{Craig:2015wwr}~(lower). All coefficients other than  $C_{\phi {\tilde{W}}}$ and $C_{\phi {\tilde{W}}B}$  are set to zero in this figure.
    }
	\label{fig:cpvdist}
\end{figure}

\subsection{CP Violation}

Understanding the source of CP violation is a goal of all future colliders. Quantum corrections to the Higgstrahlung process are sensitive to the contributions of dimension-6 CP violating operators. At tree level and to ${\cal{O}}({1/\Lambda^2})$, these operators do not interfere with the SM contributions, but at NLO, there are contributions of ${\cal{O}}({1/\Lambda^2})$ from $O_{\phi{\widetilde W}}$, $O_{\phi{\widetilde B}}$, $O_{\phi{\widetilde W}B}$ and $O_{{\widetilde W}}$ due to imaginary parts in the SM loop integrals. The CP violating contributions are odd in $\cos\theta$, where $\theta$ is the angle between the incoming electron and the outgoing $H$. The angular distributions shown at the top in Fig.~\ref{fig:cpvdist} exhibit the greatest sensitivity to  $O_{\phi{\widetilde W}}$.

We form a CP violating asymmetry to parameterize the sensitivity to each operator and summarize the results in Table~\ref{tab:cp},
\begin{align}
\label{eq:adef}
    A_{{\rm CP},i} \equiv \frac{C_i(\mu)}{\Lambda^2} \, |\Delta_i(\cos \theta < 0) - \Delta_i(\cos \theta > 0)| \, .
\end{align}
At NLO, there is no logarithmically enhanced contribution to $A_{{\rm CP},i}$.

The sensitivity to different CP violating operators is highly correlated. In the lower plot in Fig.~\ref{fig:cpvdist} we show the sensitivity to $O_{\phi{\widetilde W}B}$ and $O_{\phi{\widetilde W}}$ with all other coefficients set to zero. We assume the accuracy to be twice that of the total cross section.
Future $e^+e^-$ colliders are also sensitive to CP violation through the measurement of angular observables in the 4-body decay process $e^+e^-\rightarrow ZH\rightarrow l^+l^- b {\overline{b}}$ shown in the light green curve in Fig.~\ref{fig:cpvdist}~(lower).  It is interesting that the simple asymmetry of Eq.~\eqref{eq:adef} yields similar results as the significantly more complicated angular observables.
This is to be compared with the current limit from CP violating asymmetries in the decay $H\rightarrow 4$ leptons at the LHC~\cite{ATLAS:2023mqy}. This limit includes the quadratic contributions from the dimension-6 SMEFT operators, which accounts for the oval shape.

A strong limit on CP violation in the SMEFT comes from the electron electric dipole moment (eEDM)  measured by the ACME-II experiment~\cite{ACME:2018yjb},
\begin{equation}
    \mid d_e\mid <1.1\times 10^{-29}\,{\rm e \cdot cm}\, .
    \end{equation}
 Using tree level matching in the SMEFT~\cite{Panico:2018hal,Kley:2021yhn}, an electric dipole moment, $d_e$, corresponds to 
\begin{equation}
d_e = \sqrt{2}v \ {\rm Im}\bigg\{ \sin\theta_{\rm W} \frac{C_{eW}}{\Lambda^2}-\cos\theta_{\rm W}\frac{C_{eB}}{\Lambda^2}\bigg\} \, ,
\label{eq:edm}
\end{equation}
where $\cos\theta_{\rm W}=M_{\rm W}/M_{\rm Z}$ and coefficients $C_{eW}$ and $C_{eB}$ are evaluated at the scale $\mu=M_{\rm W}$.  The relevant CP violating coefficients,
$C_{\phi{\widetilde W}}$, $C_{\phi{\widetilde B}}$, and $C_{\phi{\widetilde W}B}$ are subsequently 
induced by renormalization group running.
The eEDM limits are shown in the lower plot in Fig.~\ref{fig:cpvdist}
and we see that the eEDM limit forces the coefficients to lie 
along a narrow line. Clearly, single parameter fits can 
be vastly misleading. The eEDM limits and the future $e^+e^-$ 
limits are complementary and will significantly constrain our understanding of CP violation in the gauge sector.

\subsection{Higgs Tri-linear and Top Quark Couplings}

\begin{figure*}
 	\centering	
    \hspace{-25pt} \includegraphics[width=.381\textwidth]{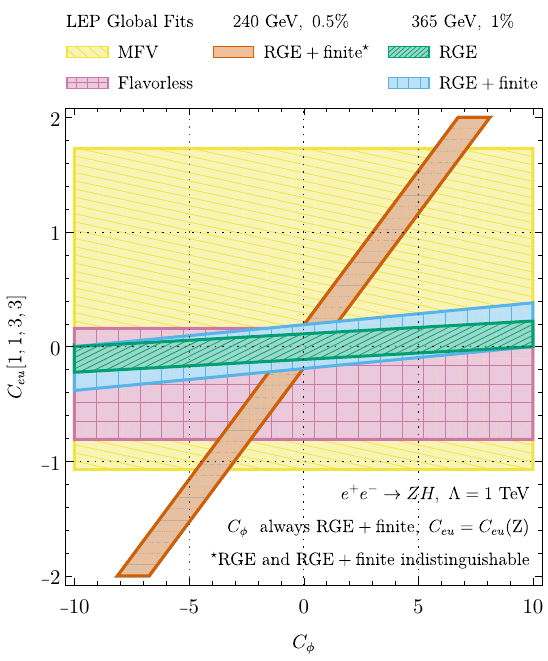}
    \hspace{40pt} \includegraphics[width=.38\textwidth]{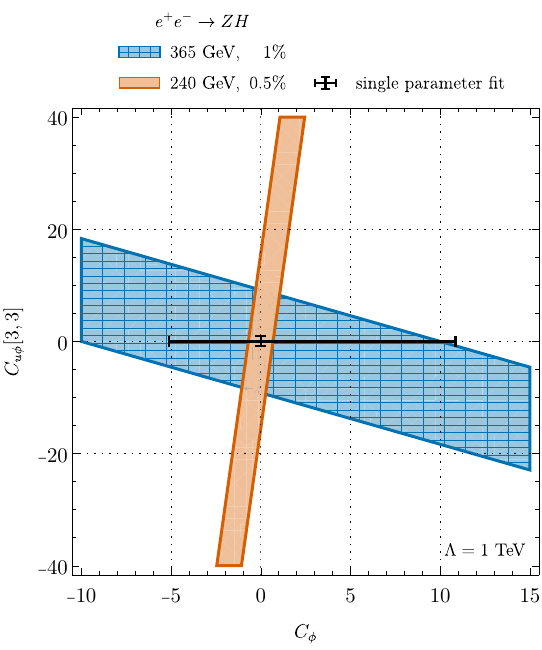} 
 	\caption{
        The  sensitivity to $C_{eu}[1,1,3,3]$ and $C_\phi$. The yellow and pink curves represent limits from global fits to $Z$ pole observables with different assumptions about the flavor structure~\cite{Bellafronte:2023amz}~(left).   
         Limits on $ C_\phi$ and $C_{u\phi}[3,3]$. The data point is the single parameter limit from a global fit to Higgs data~\cite{Ellis:2020unq} combined with $HH$ and $H$ searches~\cite{ATLAS:2024fkg}~(right). All coefficients other than those plotted are set to zero in this figure.}
 	\label{fig:ctteefig}
\end{figure*}

While the FCC-ee cannot directly produce a pair of Higgs particles, it can utilize Higgstrahlung to establish constraints on a non-Standard Model Higgs tri-linear coupling~\cite{McCullough:2013rea}, which is a result of contributions from the diagrams shown in Fig.~\ref{fig:diags}~(b).  Within the SMEFT framework, the Higgs tri-linear coupling receives contributions from the operator $O_\phi$. Having completed the  NLO calculation, we can find the correlations between the effects of all contributing operators, and here we focus on those shown in Table~\ref{tab:top}.  Of particular interest are $O_{u\phi}[3,3]$, which generates a shift in the top quark Yukawa coupling, and $O_{eu}[1,1,3,3]$ that yields the 4-fermion contributions to both $Z$ pole observables and Higgstrahlung shown in Fig.~\ref{fig:diags}~(c). 

On the left-hand side of  Fig.~\ref{fig:ctteefig}, we show the correlation between constraints on $C_\phi$ and constraints on $C_{eu}[1,1,3,3]$.  At this order, there is no logarithmic contribution to $C_\phi$, and the sensitivity curves all include the constant contribution from Table~\ref{tab:top}.  For $C_{eu}[1,1,3,3]$ we demonstrate the difference between including only the logarithms derived from the RGE and the complete calculation. The SMEFT coefficients in the plot are evaluated at the scale, $M_{\rm Z}$. At $\sqrt{s}=240~{\rm GeV}$, the finite terms are small, but at $\sqrt{s}=365~{\rm GeV}$, both the RGE and the finite contributions are equally relevant.  Combining results from the two energies will significantly constrain $C_{eu}[1,1,3,3]$. The 4-fermion operator is also constrained by $Z$-pole measurements, and the limits depend on the flavor assumptions. Assuming minimal flavor violation~\cite{DAmbrosio:2002vsn} or flavor-independent operators, $Z$ poles observables yield the 95$\%$ single parameter limits shown horizontally~\cite{Bellafronte:2023amz}.  

On the right-hand side of Fig.~\ref{fig:ctteefig},  we show the correlation between constraints on $C_\phi$ and constraints on $C_{u\phi}[3,3]$. The dependence of the Higgstrahlung rate on these coefficients is quite different at $\sqrt{s}=240~{\rm GeV}$ and $\sqrt{s}=365~{\rm GeV}$, demonstrating the opportunity offered by measurements at different energies to constrain new physics. To the order shown, there is no scale dependence of the coefficients, allowing for a direct comparison of results at various energies.
As discussed previously, these effects do not appear if one considers only an RGE analysis.

\section{Conclusion}

In this letter, we have reported on the first complete SMEFT computation at one-loop in the electroweak expansion of the Higgsstrahlung process at $e^+e^-$ colliders. We systematically studied the capabilities of a proposed Higgs factory, and specifically that of  CERN's FCC-ee, to explore BSM effects on the Higgs self-interactions, anomalous top-quark interactions, and CP violating effects that first arise at NLO and are poorly constrained by current data. We showed that the $e^+ e^-\to ZH$ process is a sensitive probe of various new physics scenarios, even when the corrections are induced by heavy new physics and enter first at 1-loop order. We demonstrated that measurements at different energies are very useful for discriminating potential scenarios and disentangling contributions due to various SMEFT operators. 

 Our plots are shown assuming a $.5\%$ accuracy at the FCC-ee on Higgstrahlung cross section measurements.  Matching the theoretical accuracy with this projected experimental accuracy  will require advances in the theoretical predictions\cite{Belloni:2022due} and a more complete understanding of the uncertainties of dimension-6 SMEFT predictions\cite{Brivio:2022pyi}.  The bottom line, however, that the interpretation of Higgstrahlung results is sensitive to the new operators that arise at NLO remains, even with a larger target accuracy.

\section*{Acknowledgements}

We are grateful to A. Freitas for helpful discussion and for providing cross-checks of the SM NLO  results.
K.A. thanks Brookhaven National Laboratory, where a significant portion of this research was conducted. 
P.P.G. is supported by the Ramón y Cajal grant~RYC2022-038517-I funded by MCIN/AEI/10.13039/501100011033 and by FSE+, and by the Spanish Research Agency (Agencia Estatal de Investigación) through the grant IFT Centro de Excelencia Severo Ochoa~No~CEX2020-001007-S. S. D. and R.S.  are supported by the U.S. Department of Energy under Grant Contract~DE-SC0012704. Digital data is provided in the ancillary file.

\bibliographystyle{apsrev4-1}
\bibliography{eezh.bib}
\end{document}